\documentclass[conference]{IEEEtran}
\IEEEoverridecommandlockouts
\ifCLASSINFOpdf
\else
\fi
\usepackage[top=0.75in, bottom=0.75in, left=0.75in, right=0.75in]{geometry}
\usepackage{bbm}
\usepackage{verbatim}
\usepackage{graphicx}
\usepackage{cite}
\usepackage{url}
\usepackage[cmex10]{amsmath}
\usepackage{amssymb}
\usepackage{amsmath}
\usepackage{algorithm, algorithmicx, algpseudocode}
\usepackage[caption=false,font=footnotesize]{subfig}
\usepackage{color}
\usepackage{cite}
\usepackage{epstopdf}
\usepackage{calc}
\usepackage{array}
\usepackage{stfloats}
\usepackage{siunitx}
\usepackage{mathtools}

\DeclareMathOperator*{\argmin}{argmin}
\DeclareMathOperator*{\argmax}{argmax}

\DeclarePairedDelimiter{\ceil}{\lceil}{\rceil}

\newcommand\norm[1]{\left\lVert#1\right\rVert}

\newcommand{\hermitian}[0]{\text{H}}
\newcommand{\transpose}[0]{\text{T}}

\begin{document}
%
\title{Tracking Sparse mmWave Channel under Time Varying Multipath Scatterers}

\author{
\IEEEauthorblockN{Veljko Boljanovic, Han Yan, and Danijela Cabric}
\IEEEauthorblockA{Electrical and Computer Engineering Department, University of California, Los Angeles\\
Email: \{vboljanovic, yhaddint\}@ucla.edu, danijela@ee.ucla.edu}
\thanks{This work is supported by NSF under grant 1718742.}
}


\IEEEspecialpapernotice{(Invited Paper)
}

\IEEEoverridecommandlockouts


\maketitle

\begin{abstract}
Due to severe signal attenuation at millimeter-wave (mmWave) frequencies large antenna arrays are required at both base station and user equipment to achieve necessary beamfoming gain and compensate for the signal power loss. The initial access and beamforming algorithms are typically designed assuming sparsity of mmWave channels, resulting from a very few significant multipath clusters, and considering fixed locations of terminals and scatterers. Channel tracking algorithms have been proposed to account for channel variations due to user mobility. Existing works did not consider mobility of the scatterers, which adds new challenges and opportunities into a channel tracking problem. In this work, we consider a more realistic assumption of mobile scatterers and their impact on channel tracking algorithms. We propose a novel channel tracking algorithm that takes into account the dynamics of cluster evolution, and adaptively tracks channel parameters with the objective to reduce training overhead. We also propose a simple implementation of aperiodic tracking to accommodate tracking to different channel variations. We analyze the performance of the proposed tracking algorithm under highly dynamic channels, and compare it to existing channel tracking algorithms with respect to tracking accuracy, achievable rate, and required training overhead, when aperiodic and periodic trackings are used.
\end{abstract}


%
\IEEEpeerreviewmaketitle
%
%
\section{Introduction}
\label{sec:introduction}
Communication over millimeter-wave (mmWave) frequency bands is considered as one of the key features in the fifth generation New Radio (5G-NR) standard  \cite{5GNR_rel15}. Although mmWave frequencies offer abundant spectrum and hence high data rates, their use comes at a cost of less favorable propagation conditions \cite{Rappaport:mmWavewillwork}. Both the transmitter and the receiver are required to use large antenna arrays with up-to-date channel state information (CSI) to achieve the beamforming gain and to compensate for severe propagation loss. Conventionally, the CSI is obtained through periodic channel estimation, whereas a number of advanced channel tracking techniques have been proposed recently to reduce inevitable training overhead. It is commonly assumed that mmWave wireless channel is sparse, meaning that there are a few significant signal paths coming to the receiver. Thus, the problem of acquiring the CSI for narrowband communications reduces to estimation and tracking of few parameters that describe those paths - angles of departure (AoD), angles of arrival (AoA), and path gains. Previous works on channel tracking can be roughly divided into two groups based on assumed spatial consistency model in the channel. The first group does not consider channel geometry and often assumes that AoD, AoD, and path gains change according to the certain statistical models, e.g., Gauss-Markov processes, while the other line of works takes geometry into account and models changes of channel parameters accordingly.

A number of papers on channel tracking assuming statistical models was published recently \cite{He:mmwavetracking, Va:beamtracking, Jayaprakasam:robust}. In \cite{He:mmwavetracking}, authors apply rotation matrices onto old beamforming vectors to create new ones for tracking. Moreover, \cite{He:mmwavetracking} considers hybrid analog-digital beamforming and reduces overhead by using multiple RF chains for channel sounding. In \cite{Va:beamtracking, Jayaprakasam:robust}, channel tracking using Kalman filter is investigated. This approach provides continuous channel tracking, but it is sensitive to noise and estimation error accumulates fast. Recent works also consider channel geometry and model parameter changes in a deterministic way \cite{Yan:wideband, Palacios:tracking}. The work \cite{Yan:wideband} studies wideband channel tracking using geometry-based spatial channel model (GSCM) which assumes static scatterers and constant angles of departure \cite{5G:mmMagic}, while \cite{Palacios:tracking} uses a ray-tracing tool to accurately model mobility effects and spatial consistency. 

Recent mmWave measurement campaign in \cite{Bas:mobilescatter} showed that a significant part of received signal energy comes as a reflection from mobile objects in the channel, including cars and pedestrians, hence, these objects can be considered as mobile scatterers. In urban environment with densely deployed base stations, distances between the base station, the mobile stations, and scatterers are small, so mobile scatterers can cause significant AoD/AoA changes even for static users. Due to fairly predictive movements of scatterers without sudden direction changes, AoD and AoA will tend to change in one direction with high probability. However, AoD and AoA can experience changes in the other direction due to unexpected array orientation changes, e.g., due to rotations. In this work, we propose a new channel tracking algorithm that takes advantage of predictive AoD/AoA changes in GSCM and remains robust to potential changes in the other direction. Similar to \cite{Marzi:CStracking}, the proposed algorithm initially uses compressed sensing (CS) approach with pseudo-random sequences for beamformer precoding and combining, but later it exploits knowledge of the previous parameter estimates to further reduce channel tracking overhead. Unlike beam tracking in the Third Generation Partnership Project New Radio (3GPP-NR) Standard \cite{Zorzi:tutorial}, where reduction in overhead is achieved by steering certain number of beams around the previous estimates, the proposed algorithm achieves overhead reduction by using projections of pseudo-random sequences around the previous estimates. We also propose a simple implementation of aperiodic channel tracking, and use it in conjunction with the proposed algorithm and algorithms from \cite{Marzi:CStracking} and \cite{ Zorzi:tutorial} to compare their AoD/AoA tracking accuracy, achievable rate with aperiodic channel tracking, and overhead when aperiodic channel tracking is used. In addition, we compare these algorithms in terms of achievable rate when periodic tracking with maximum overhead is used.

\begin{figure}
\begin{center}
\includegraphics[width=0.49\textwidth]{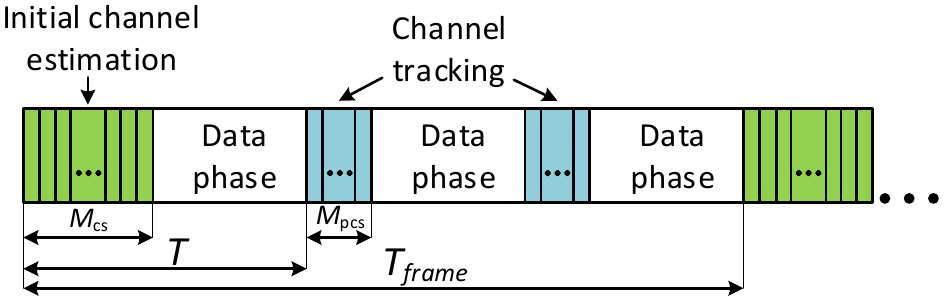}
\end{center}
\vspace{-4mm}
\caption{The frame structure used in this work.}
\vspace{-4mm}
\label{fig:frame}
\end{figure}

The rest of the paper is organized as follows. In Section~\ref{sec:models}, we introduce the channel and system models. Section~\ref{sec:proposed} describes the proposed algorithm and a simple implementation of aperiodic tracking. In Section~\ref{sec:evaluation}, we compare the proposed algorithm with existing channel tracking algorithms. Finally, conclusions are drawn in Section~\ref{sec:conclusions}.

\textit{Notations:} Scalars, vectors, and matrices are denoted by non-bold, bold lower-case, and bold upper-case letters, respectively, e.g., $h$, $\mathbf{h}$ and $\mathbf{H}$. Transpose, conjugate transpose, and pseudo-inverse are denoted by $(.)^{\transpose}$, $(.)^{\hermitian}$, $(.)^+$, respectively. The $l_2$-norm of a vector $\mathbf{h}$ is denoted by $||\mathbf{h}||$. Operator $\text{diag}(\mathbf{A})$ aligns diagonal elements of $\mathbf{A}$ in a vector.

%
%
\section{Channel and System Models}
\label{sec:models}

We consider a narrowband mmWave MIMO system with uniform linear arrays (ULA) consisting of $N_{\text{BS}}$ and $N_{\text{MS}}$ antenna elements at the base station (BS) and the mobile station (MS), respectively. There are $L$ distinct single-ray signal paths that are coming to the mobile station in the downlink (DL). We focus on the azimuth plane, and the channel matrix $\mathbf{H}$ is
\begin{align}
\mathbf{H} = \sqrt{\frac{N_{\text{BS}}N_{\text{MS}}}{L}}\sum_{l=1}^{L}g_l \mathbf{a}_{\text{MS}}(\phi_l)\mathbf{a}_{\text{BS}}^{\text{H}}(\theta_l)
\label{eq:channel_model}
\end{align}
where $g_l = e^{j\psi_l}$, $\theta_l \in [-\frac{\pi}{2}, \frac{\pi}{2}]$, and $\phi_l \in [-\frac{\pi}{2}, \frac{\pi}{2}]$, represent the complex path gains, AoDs, and AoAs, respectively. Phase $\psi_l$ in the complex gain is a uniform random variable in the interval $[-\pi,\pi)$. The vectors $\mathbf{a}_{\text{BS}}(\theta_l) \in \mathbb{C}^{N_{\text{BS}}}$ and $\mathbf{a}_{\text{MS}}(\phi_l) \in \mathbb{C}^{N_{\text{MS}}}$ are the spatial responses corresponding to the BS and the MS, and they are defined as
\begin{align}
\mathbf{a}_{\text{BS}}(\theta_l) = \frac{[1,e^{-j\pi\sin(\theta_l)},...,e^{-j(N_{\text{BS}}-1)\pi\sin(\theta_l)}]^{\text{T}}}{\sqrt{N_{\text{BS}}}},
\label{eq:spatialAoD}
\end{align}
\begin{align}
\mathbf{a}_{\text{MS}}(\phi_l) = \frac{[1,e^{-j\pi\sin(\phi_l)},...,e^{-j(N_{\text{MS}}-1)\pi\sin(\phi_l)}]^{\text{T}}}{\sqrt{N_{\text{MS}}}}.
\label{eq:spatialAoA}
\end{align}
For the sake of simplicity, we can assume that $L=1$, i.e., there is one Non-Line-of-Sight (NLoS) path coming to the receiver from a mobile scatterer. Therefore, we can remove summation and index $l$ from (\ref{eq:channel_model}) and use notation $\mathbf{H}(g, \theta, \phi)$ to indicate that $\mathbf{H}$ depends on parameters $g$, $\theta$, and $\phi$. 

We consider DL communication with the frame structure depicted in Fig.~\ref{fig:frame}. Each time frame is of length $T_{frame}$ and it consists of 10000 slots. At the beginning of the frame, channel parameter estimates $\hat{g}$, $\hat{\theta}$, and $\hat{\phi}$ are acquired through the initial channel estimation, fed back to the BS during a short period reserved for control messages\footnote{We assume non-standalone (NSA) mmWave system architecture with Long Term Evolution (LTE) link between the BS and the MS \cite{Giordani:NSA} used for control information exchange for a short period of time after channel estimation/tracking.}, and then used for beam steering in data communication phase. It is assumed that the channel does not change during the initial channel estimation and tracking periods. However, due to high user and scatterer mobilities, channel parameters $g$, $\theta$, and $\phi$ can significantly change during the data communication phase, therefore, beam tracking and data communication slots must be interleaved, as depicted in Fig.~\ref{fig:frame}, to maintain the required link budget. After each channel tracking period, channel parameter estimates are fed back to the BS, and beamsteering vectors are updated for the following data phase. Channel tracking can be done in a periodic or aperiodic manner \cite{3GPP:per_aper_track}, i.e., period $T$ can be fixed or time-varying within the frame. In this work, we propose and explain a simple implementation of aperiodic tracking in Sec.~\ref{sec:aperiodic}.

We focus on phase-only analog beamforming where both the BS and the MS have only one RF chain, and they can steer beams in  $Q_{\text{BS}}$ and $Q_{\text{MS}}$ angles from sets $\mathcal{A}_{\text{BS}}=\{-\frac{\pi}{2}+(Q_{\text{BS}}-i) \frac{\pi}{Q_{\text{BS}}}~|~i=1,2,...,Q_{\text{BS}}\}$ and $\mathcal{A}_{\text{MS}}=\{-\frac{\pi}{2}+(Q_{\text{MS}}-i) \frac{\pi}{Q_{\text{MS}}}~|~i=1,2,...,Q_{\text{MS}}\}$, respectively. At $k$-th time slot, the BS uses precoding vector $\mathbf{f}_k \in \mathbb{C}^{N_{\text{BS}}}$ and the MS uses combining vector $\mathbf{w}_k \in \mathbb{C}^{N_{\text{MS}}}$ for beam steering. Elements of $\mathbf{f}_k$ have magnitudes $\frac{1}{\sqrt{N_{\text{BS}}}}$, and elements of $\mathbf{w}_k$ have magnitudes $\frac{1}{\sqrt{N_{\text{MS}}}}$. Note that scaling factor $\sqrt{N_{\text{BS}}N_{\text{MS}}}$ in (\ref{eq:channel_model}) compensates for unit beamforming gains while the channel still remains normalized. Assuming perfect synchronization and unit pilot symbol, the received signal at the $k$-th slot can be expressed as
\begin{align}
y_k = \mathbf{w}_k^{\hermitian} \mathbf{H} \mathbf{f}_k + n_k,
\label{eq:received_signal}
\end{align}
where $n_k \in \mathcal{CN}(0,\sigma_n^2)$ is post-beamforming noise.

%
%
\section{Proposed Tracking Algorithm}
\label{sec:proposed}
We propose a two-step tracking algorithm based on pseudo-random precoding and combining vectors. The first step is the initial channel estimation based on CS approach \cite{Marzi:CStracking}, while the second step includes a novel channel tracking algorithm that utilizes information about previous estimates of channel parameters to further reduce overhead.

\subsection{Initial Channel Estimation}
\label{sec:initial_training}

\begin{figure}
\begin{center}
\includegraphics[width=0.49\textwidth]{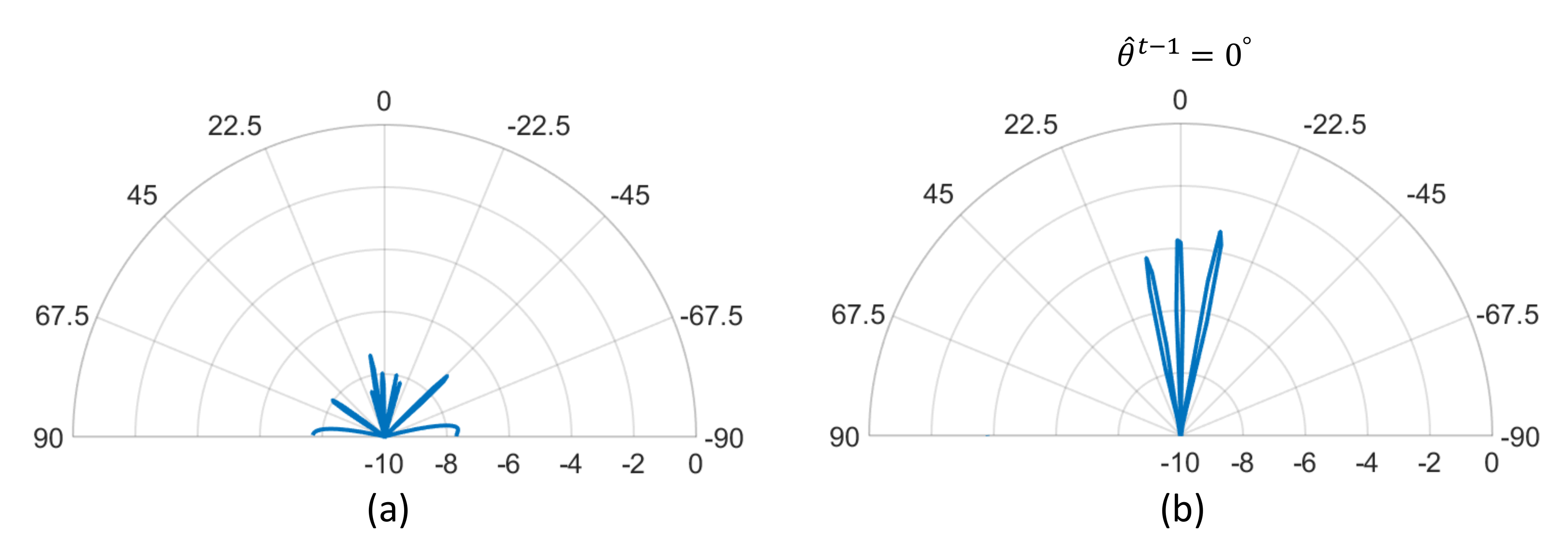}
\end{center}
\vspace{-4mm}
\caption{Beam pattern comparison. (a) Normalized quasi-omnidirectional beam pattern generated by a random vector. (b) Normalized beam pattern of projection of a random vector onto column space of $\mathbf{F}_1$ with $\hat{\theta}^{t-1}=\ang{0}$}
\vspace{-4mm}
\label{fig:patterns}
\end{figure}

We exploit the sparsity of mmWave channel and use CS approach for initial parameter estimation. In the $k$-th slot of the initial channel estimation, the BS uses precoding vector $\mathbf{f}_k$ and the MS uses combining vector $\mathbf{w}_k$, where elements of both vectors are drawn randomly from the set $\{\pm 1 \pm j\}$. For normalization purposes, vectors $\mathbf{f}_k$ and $\mathbf{w}_k$ are multiplied by $\frac{1}{\sqrt{2N_{\text{BS}}}}$ and $\frac{1}{\sqrt{2N_{\text{MS}}}}$, respectively. An example of quasi-omnidirectional beam pattern generated by $\mathbf{f}_k$ or $\mathbf{w}_k$ is presented in Fig.~\ref{fig:patterns}(a). By using superscript $t$ in $g^t$, $\theta^t$, and $\phi^t$ to denote estimation/tracking period, the received signal after $M_{\text{cs}}$ slots (measurements) at $t=1$ is
\begin{align}
\mathbf{y} = \text{diag}(\mathbf{W}^{\hermitian}_{\text{cs}} \mathbf{H}(g^1,\theta^1,\phi^1) \mathbf{F}_{\text{cs}}) + \mathbf{n},
\label{eq:received_cs}
\end{align}
where $\mathbf{F}_{\text{cs}} = [\mathbf{f}_1,...,\mathbf{f}_{M_{\text{cs}}}]$, $\mathbf{W}_{\text{cs}} = [\mathbf{w}_1,...,\mathbf{w}_{M_{\text{cs}}}]$, and $\mathbf{n} \in \mathbb{C}^{M_{\text{cs}}}$ is a Gaussian random vector with $\mathcal{CN}(\mathbf{0},\sigma_n^2 \mathbf{I}_{M_{\text{cs}}})$. By defining $\mathbf{z}(\theta, \phi) = \text{diag}(\mathbf{W}^{\hermitian} \mathbf{H}(\theta,\phi) \mathbf{F})$, the expression becomes
\begin{align}
\mathbf{y} = g^1\mathbf{z}(\theta^1, \phi^1) + \mathbf{n}.
\label{eq:received_cs_shorter}
\end{align}
The maximum likelihood (ML) estimates of parameters can be found as
\begin{align}
\hat{g}^1,\hat{\theta}^1,\hat{\phi}^1 = \argmin_{g,\theta,\phi} \norm{\mathbf{y} - g\mathbf{z}(\theta,\phi)}^2.
\label{eq:ml_estimates}
\end{align}
Using least squares (LS), we estimate the path gain $g$ as
\begin{align}
\hat{g}^1 = \frac{\mathbf{y}^{\hermitian}\mathbf{z}(\theta,\phi)}{\norm{\mathbf{z}(\theta, \phi)}^2},
\label{eq:ls_estimate}
\end{align}
for any angles $\theta$ and $\phi$. If (\ref{eq:ls_estimate}) is substituted in (\ref{eq:ml_estimates}), the ML angle estimates are calculated as
\begin{align}
\hat{\theta}^1,\hat{\phi}^1 = \argmax_{\theta \in \mathcal{A}_{\text{BS}},\phi \in \mathcal{A}_{\text{MS}}}\frac{\mathbf{y}^{\hermitian}\mathbf{z}(\theta,\phi)}{\norm{\mathbf{z}(\theta, \phi)}^2}.
\label{eq:ml_angles}
\end{align}
The ML estimation in (\ref{eq:ml_angles}) assumes that the MS have knowledge of $\mathbf{F}_{\text{cs}}$, which is obtained just once during the first control information exchange and then reused in all time frames. The total number of measurements $M_{\text{cs}}$ needed for reliable parameter estimation in the CS scales linearly with the number of paths $L$ \cite{Marzi:CStracking}. In this work, we assume that estimation is reliable if the root mean square error (RMSE) of AoD/AoA estimation vanishes at $0$dB signal-to-noise ratio (SNR), and our simulations show that $M_{\text{cs}}=45$ is enough for reliable estimation when $N_{\text{BS}} = N_{\text{MS}} = 32$.

\begin{figure}
\begin{center}
\includegraphics[width=0.49\textwidth]{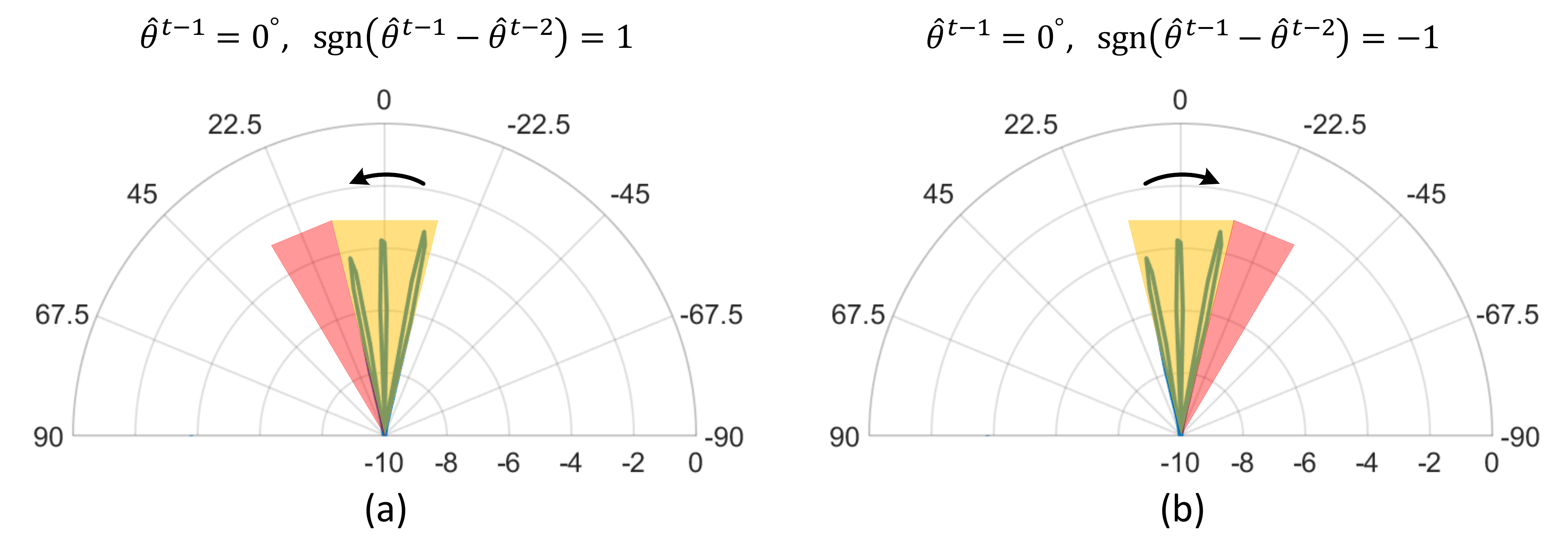}
\end{center}
\vspace{-4mm}
\caption{Scanned beamspace. (a) Scanned beamspace at the BS with $\text{sgn}(\hat{\theta}^{t-1} - \hat{\theta}^{t-2})=1$, i.e., with counter-clockwise angular change. (b) Scanned beamspace at the BS with $\text{sgn}(\hat{\theta}^{t-1} - \hat{\theta}^{t-2})=-1$, i.e., with clockwise angular change.}
\vspace{-4mm}
\label{fig:beamspace}
\end{figure}

\subsection{Channel Tracking: Projected Compressed Sensing (PCS)}
\label{sec:proposed_tracking}
The proposed algorithm relies on the previous angle estimates $\hat{\theta}^{t-1}$ and $\hat{\phi}^{t-1}$ from tracking period $t-1$ to track channel parameters because the current angles $\theta^t$ and $\phi^t$ are close to the previous ones with high probability. Since $\hat{\theta}^{t-1}$ and $\hat{\phi}^{t-1}$ are known, there is no need to use quasi-omnidirectional beam patterns for tracking, and beamspace around $\hat{\theta}^{t-1}$ and $\hat{\phi}^{t-1}$ can be scanned instead. The tracking procedures at the BS and the MS are identical, and we describe just the former for the sake of brevity.

We first take a pseudo-random vector $\mathbf{f}$ from the initial channel estimation. Using the previous estimate $\hat{\theta}^{t-1}$ we create four matrices $\mathbf{F_1}$, $\mathbf{F_2}$, $\mathbf{F_3}$, and $\mathbf{F_4}$, in the way described in (\ref{eq:F1}) and (\ref{eq:F2}). The angular shift of $\frac{2 \pi}{N_{\text{BS}}}$ makes the columns of these matrices approximately orthogonal, since for large antenna arrays $\mathbf{a}_{\text{BS}}(\alpha)^{\hermitian} \mathbf{a}_{\text{BS}}(\alpha + \frac{2 \pi}{N_{\text{BS}}}) \approx 0$ for any $\alpha \in [-\pi,\pi)$. The additional angular shifts for $\mathbf{F}_2$, $\mathbf{F}_3$, and $\mathbf{F}_4$, are defined as $\delta_{\mathbf{F}_2} = \frac{1}{4}\text{sgn}(\hat{\theta}^{t-1} - \hat{\theta}^{t-2}) \frac{2 \pi}{N_{\text{BS}}}$, $\delta_{\mathbf{F}_3} = \frac{1}{2}\text{sgn}(\hat{\theta}^{t-1} - \hat{\theta}^{t-2}) \frac{2 \pi}{N_{\text{BS}}}$, and $\delta_{\mathbf{F}_4} = \frac{3}{4}\text{sgn}(\hat{\theta}^{t-1} - \hat{\theta}^{t-2}) \frac{2 \pi}{N_{\text{BS}}}$, respectively. The function $\text{sgn}(x)$ is a sign function of $x$, and it takes value $1$ if $x \ge 0$, and $-1$ otherwise.

\begin{figure*}[t]
\begin{align}
\mathbf{F_1} = \left[ \mathbf{a}_{\text{BS}}(\hat{\theta}^{t-1}), \mathbf{a}_{\text{BS}} \left( \hat{\theta}^{t-1}+\frac{2 \pi}{N_{\text{BS}}} \right), \mathbf{a}_{\text{BS}} \left( \hat{\theta}^{t-1}-\frac{2 \pi}{N_{\text{BS}}} \right) \right]
\label{eq:F1}
\end{align}
\begin{align}
\mathbf{F}_i = \left[ \mathbf{a}_{\text{BS}} (\hat{\theta}^{t-1} + \delta_{\mathbf{F}_i}), \mathbf{a}_{\text{BS}}\left( \hat{\theta}^{t-1}+\frac{2 \pi}{N_{\text{BS}}} + \delta_{\mathbf{F}_i} \right), \mathbf{a}_{\text{BS}}\left( \hat{\theta}^{t-1}-\frac{2 \pi}{N_{\text{BS}}} + \delta_{\mathbf{F}_i} \right) \right],~~i=2,3,4
\label{eq:F2}
\end{align}
\vspace{-5mm}
\end{figure*}

The vector $\mathbf{f}$ is projected onto column spaces of all four matrices, i.e. four new vectors are obtained as follows
\begin{align}
\mathbf{f}_1 = \mathbf{F}_1\mathbf{F}_1^+\mathbf{f},~\mathbf{f}_2 = \mathbf{F}_2\mathbf{F}_2^+\mathbf{f},~\mathbf{f}_3 = \mathbf{F}_3\mathbf{F}_3^+\mathbf{f},~\mathbf{f}_4 = \mathbf{F}_4\mathbf{F}_4^+\mathbf{f}.
\end{align}
Since we focus on phase-only analog beamforming, magnitude of each element in $\mathbf{f}_1$, $\mathbf{f}_2$, $\mathbf{f}_3$, and $\mathbf{f}_4$ must be scaled to $\frac{1}{\sqrt{N_{\text{BS}}}}$, which also ensures that all vectors have unit norm. An example of the beam pattern of $\mathbf{f}_1$ is presented Fig.~\ref{fig:patterns}(b). Due to nearly orthogonal columns of $\mathbf{F}_1$, most of the signal energy is in three distinct beam lobes. The beam patterns for $\mathbf{f}_2$, $\mathbf{f}_3$, and $\mathbf{f}_4$, look similar, but are rotated for $\delta_{\mathbf{F}_2}$, $\delta_{\mathbf{F}_3}$, and $\delta_{\mathbf{F}_4}$, respectively. Note that these rotations depend on $\text{sgn}(x)$ function, i.e., on estimated direction of angular changes based on the last two tracking periods. Illustrations of scanned beamspace with $\text{sgn}(\hat{\theta}^{t-1} - \hat{\theta}^{t-2})=1$ and $\text{sgn}(\hat{\theta}^{t-1} - \hat{\theta}^{t-2})=-1$ are provided in Fig.~\ref{fig:beamspace}(a) and Fig.~\ref{fig:beamspace}(b). The rotations provide asymmetric scanning around the last estimate $\hat{\theta}^{t-1}$ by ensuring that an additional part of beam space in the estimated direction of angular change (the part colored in red) is scanned.

The precoding matrix for $4$ slots (measurements) is constructed as $\mathbf{F}_{\text{pcs}} = [\mathbf{f}_1, \mathbf{f}_2, \mathbf{f}_3, \mathbf{f}_4]$. After a similar procedure, the matrix $\mathbf{W}_{\text{pcs}}$ is obtained, and the received signal at $t$-th estimation/tracking period is
\begin{align}
\mathbf{y} = \text{diag}(\mathbf{W}^{\hermitian}_{\text{pcs}} \mathbf{H}(g^t,\theta^t,\phi^t) \mathbf{F}_{\text{pcs}}) + \mathbf{n},
\end{align}
where $\mathbf{n} \in \mathcal{CN}(\mathbf{0}, \sigma_n^2\mathbf{I}_4)$. However, due to pseudo-randomness in $\mathbf{f}$ and imperfect orthogonality among columns of $\mathbf{F}_i, i=1,2,3,4$, beam patterns of projections can have less than three distinct lobes, which causes gaps in the scanned beamspace. In addition, most of the signal energy is distributed in up to three directions and full beamforming gain is not achieved. Therefore, more than one pseudo-random sequence must be projected to increase diversity. Our simulations show that projections of five sequences are enough to achieve vanishing AoD/AoA RMSE at $\text{SNR}=0$dB. In other words, reliable estimation requires precoding matrix $\mathbf{F}_{\text{pcs}} \in \mathbb{C}^{N_{\text{BS}}\times M_{\text{pcs}}}$ with $M_{\text{pcs}}=20$ precoding vectors, i.e., $\mathbf{F}_{\text{pcs}} = \left[\mathbf{f}^1_1, \mathbf{f}^1_2, \mathbf{f}^1_3, \mathbf{f}^1_4, \mathbf{f}^2_1, \mathbf{f}^2_2, \mathbf{f}^2_3, \mathbf{f}^2_4, ..., \mathbf{f}^5_1, \mathbf{f}^5_2, \mathbf{f}^5_3, \mathbf{f}^5_4\right]$, where the superscripts denote pseudo-random sequences and the subscripts denote projections. Finally, the channel parameter estimates $\hat{g}^t$, $\hat{\theta}^t$, and $\hat{\phi}^t$, can be found from the received signal $\mathbf{y} \in \mathbb{C}^{M_{\text{pcs}}}$ using the ML approach described in (\ref{eq:received_cs_shorter})-(\ref{eq:ml_angles}). Note that the BS does not have to share information about $\mathbf{F}_{\text{pcs}}$ used in the next tracking period with the MS, since $\mathbf{F}_{\text{pcs}}$ can be created using the AoD estimate $\hat{\theta}^t$ and five out of $M_{\text{cs}}$ pseudo-random precoding sequences used in the initial channel estimation.

\subsection{Implementation of Aperiodic Tracking}
\label{sec:aperiodic}
The 3GPP-NR Standard supports the following set of possible tracking periodicities: $S_T=\{70, 140, 280, 560, 1120, 2240, 4480, 8960\}$ slots\footnote{We scale periodicities by 14 since the notion of slot in \cite{Zorzi:tutorial} is different and it has 14 symbols.} \cite{Zorzi:tutorial}. In this work, we propose a simple technique for aperiodic tracking where period $T$ does not remain fixed over a period of one frame, but it rather changes adaptively by taking different values from $S_T$.
We first define the maximum AoD/AoA change $\gamma_{\text{max}}$ that can be tolerated as a half of $3$dB beamwidth. For example, for large antenna arrays with $32$ elements, this value is approximately $\gamma_{\text{max}}= \ang{2.5}$. After the initial beam training, the first period $T^1$ should take smaller values from $S_T$, e.g., one of the first four values, to avoid channel changes much greater than $\gamma_{\text{max}}$. In each following tracking instance, period $T^t, t=2,3...$ is computed according to the ratio between $\gamma_{\text{max}}$ and the maximum estimated angular change $\Delta^t$
\begin{align}
T^t_{\mathbb{Z}^+} = \ceil[\bigg]{\frac{\gamma_{\text{max}}}{\Delta^t}T^{t-1}}
\label{eq:T_aperiodic}
\end{align}
where $\Delta^t = \text{max} \{ \lvert \hat{\theta}^t - \hat{\theta}^{t-1} \rvert, \lvert  \hat{\phi}^t - \hat{\phi}^{t-1} \rvert \}$. The function $\ceil{x}$ rounds $x$ up to the closest integer. Since $T^t_{\mathbb{Z}^+}$ in (\ref{eq:T_aperiodic}) can be any positive interger, it has to be quantized to one of the values from $S_T$. We define a range around each value from $S_T$ using the midpoints between adjacent values, and we quantize $T^t_{\mathbb{Z}^+}$ in the following way
\begin{align}
T^t=\begin{cases}
    70, & \text{if $1 \le T^t_{\mathbb{Z}^+} < 105$}\\
    140, & \text{if $105 \le T^t_{\mathbb{Z}^+} < 210$}\\
    ~... & \text{}\\
    8960, & \text{if $6720 \le T^t_{\mathbb{Z}^+}$}.\\
\end{cases}
\label{eq:quatization}
\end{align}
If the previous estimates of AoD and AoA are the same as their current estimates, i.e., if $\Delta^t = 0$, then $T^t = 2T^{t-1}$ assuming that $T^{t-1} < 8960$. Information about the current period $T^t$ is fed back to the BS as a part of control information exchange. Hence, both the BS and the MS know when the next tracking is going to be triggered.

%
%
\section{Numerical Evaluation}
\label{sec:evaluation}

In this section, we numerically evaluate the proposed channel tracking algorithm and compare it to two existing channel tracking algorithms including beam-sweeping from 3GPP-NR Standard \cite{Zorzi:tutorial} and compressed sensing based tracking \cite{Marzi:CStracking}. Although the 3GPP-NR Standard sets the maximum number of beam pairs for beam-sweeping to 64, the initial channel estimation with large antenna arrays and analog beamforming requires even higher number of beam pairs due to small beam width\footnote{The previous works usually assumed $N_{\text{BS}}N_{\text{MS}}$ beam pairs.}. We exclude the initial channel estimation and compare these algorithms based on their channel tracking performance in order to make required overheads comparable.

The 3GPP-NR Standard does not specify the maximum number of directions that is scanned at the BS/MS during the channel tracking, and in \cite{Zorzi:tutorial} the authors consider 4 directions around the previous angle estimate as an option. However, our simulations show that the tracking performance is significantly boosted if an additional direction along the previous estimate is scanned, thus, we use $M_{\text{bs}}=25$ beam pairs\footnote{The number of beam pairs is equal to the number of slots in this work.} for beam-sweeping with resolution of 32 possible angles at both the BS and the MS. As discussed in Sec.~\ref{sec:proposed}, for vanishing AoD/AoA RMSE at $\text{SNR}=0$dB, the compressed sensing approach and the projected compressed sensing require $M_{\text{cs}}=45$ and $M_{\text{pcs}}=20$ slots, respectively.

\begin{figure}[t]
\begin{tabular}{cc}
\vspace{0mm}
\subfloat[The first model of AoD/AoA changes.]{%
  \includegraphics[clip,width=0.98\columnwidth]{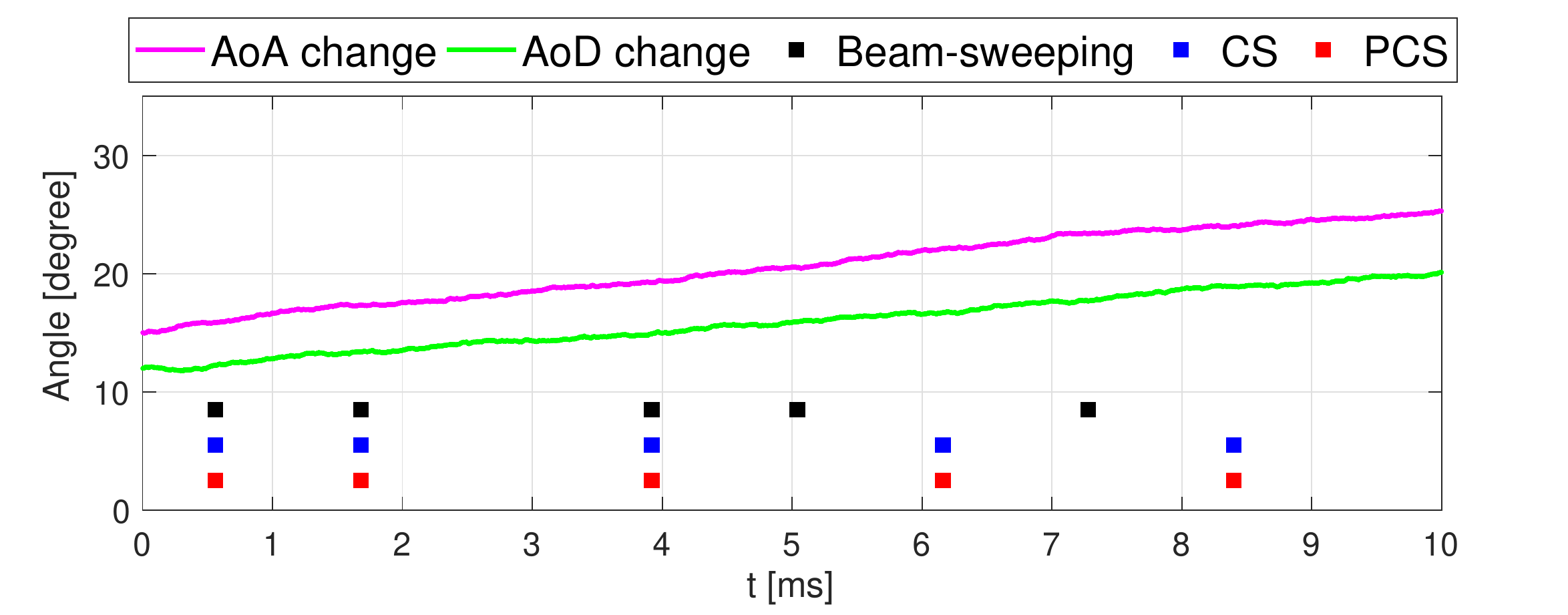}%
}\\
\vspace{-3mm}
\subfloat[The second model of AoD/AoA changes.]{%
  \includegraphics[clip,width=0.98\columnwidth]{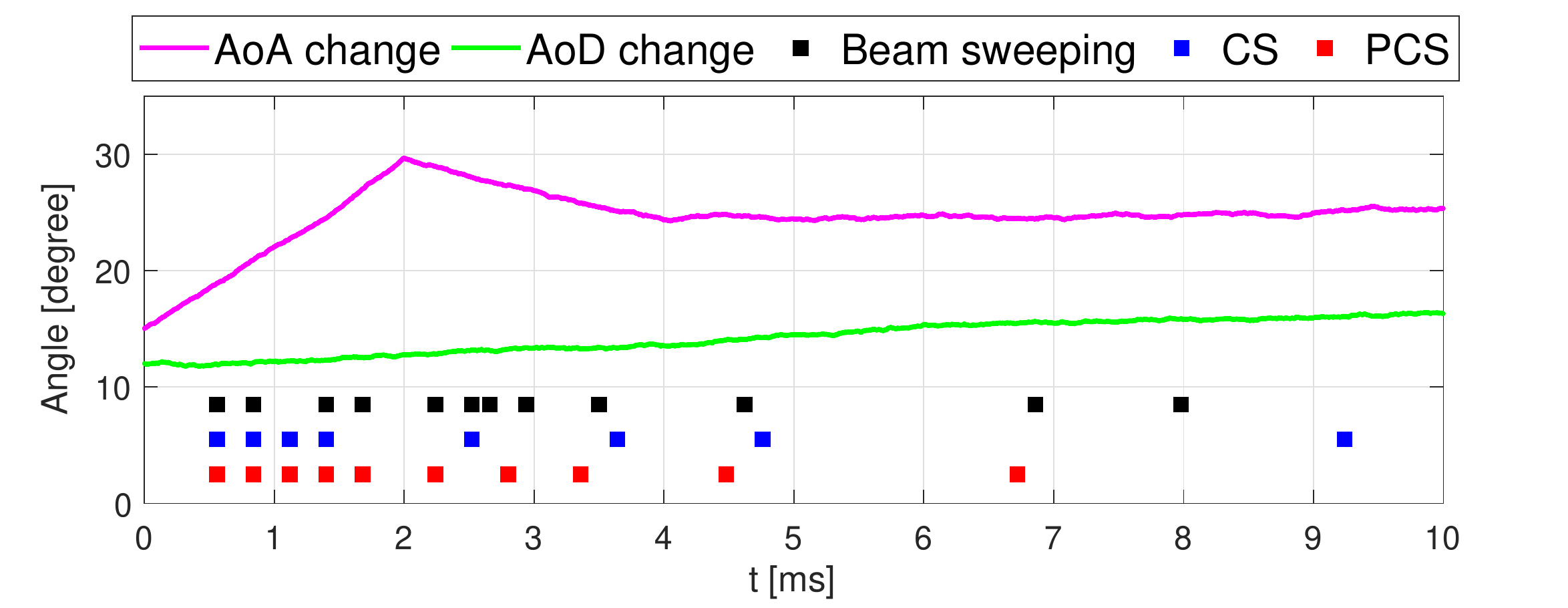}%
}\\
\vspace{0mm}
\end{tabular}
\caption{The AoD/AoA change models used in this work, along with time instances when algorithms perform aperiodic tracking.}
\vspace{-5mm}
\label{fig:triggers}
\end{figure}

We consider a communication system with $N_{\text{BS}}=32$ and $N_{\text{MS}}=32$ antennas at the BS and the MS, respectively. Both the BS and the MS have angular resolutions of $Q_{\text{BS}}=Q_{\text{MS}}=256$ angles. Since there is $L=1$ path, we assume without loss of generality that the path gain $g=e^{j \psi}$ preserves unit magnitude and randomly changes the phase $\psi$ during the data communication phase. We consider two piece-wise linear models of AoD/AoA changes, and we extend them by adding Gaussian random variables in both models. In the first model, AoD $\theta^k$ and AoA $\phi^k$ at $k$-th slot are given by
\begin{align}
\theta^k = \theta^{k-1} + \frac{\Delta \theta}{10000} + \Theta,
\label{eq:AoDchanges}\\
\phi^k = \phi^{k-1} + \frac{\Delta \phi}{10000} + \Phi,
\label{eq:AoAchanges}
\end{align}
where $\Delta \theta=\ang{10}$, $\Delta \phi=\ang{10}$, $\Theta \in \mathcal{N}(0,10^{-4})$, and $\Phi \in \mathcal{N}(0,10^{-4})$. In the second model, the angles at $k$-th slot are described as in (\ref{eq:AoDchanges})-(\ref{eq:AoAchanges}), but with $\Delta \theta=\ang{5}$ and  $\Delta \phi$ defined as
\begin{align}
\Delta \phi = \begin{cases}
                \ang{15}, & \text{if $1 \le k < 2000$}\\
                -\ang{5}, & \text{if $2000 \le k < 4000$}\\
                 \ang{1}, & \text{if $4000 \le k \le 10000$}.\\
                \end{cases}
\end{align}

Note that we allow AoD/AoA changes across all time slots just to ensure equal angle changes for all algorithms when aperiodic tracking is used. When channel tracking is performed, angles are assume to be fixed, i.e., we use $\theta^{k-1}$ and $\phi^{k-1}$ corresponding to the last data communication slot.

Described AoD/AoA change models are depicted in Fig.~\ref{fig:triggers} along with time instances when different algorithms trigger aperiodic tracking at $\text{SNR}=0$dB. We set $T^1=560$ and $\gamma_{\text{max}}=\ang{2.5}$ slots for all algorithms, and then adaptively change $T^t$ for $t=2,3,...$, as described in Sec.~\ref{sec:aperiodic}. For the first model, period $T^t$ should remain constant within the frame due to nearly linear angle changes. This happens for the CS-based and PCS-based algorithms with high probability, while beam-sweeping fails to properly trigger channel tracking. Due to randomness in the model, channel tracking time instances can be offset or differently distributed within the frame for all three algorithms. The second model requires frequent channel tracking for fast changes, and the frequency reduces as the angle changes become slower. Again, the CS-based and the PCS-based algorithms perform better than beam-sweeping, but their channel tracking time instances can slightly vary due to randomness.

\begin{figure}[t]
\begin{center}
\includegraphics[clip,width=0.46\textwidth]{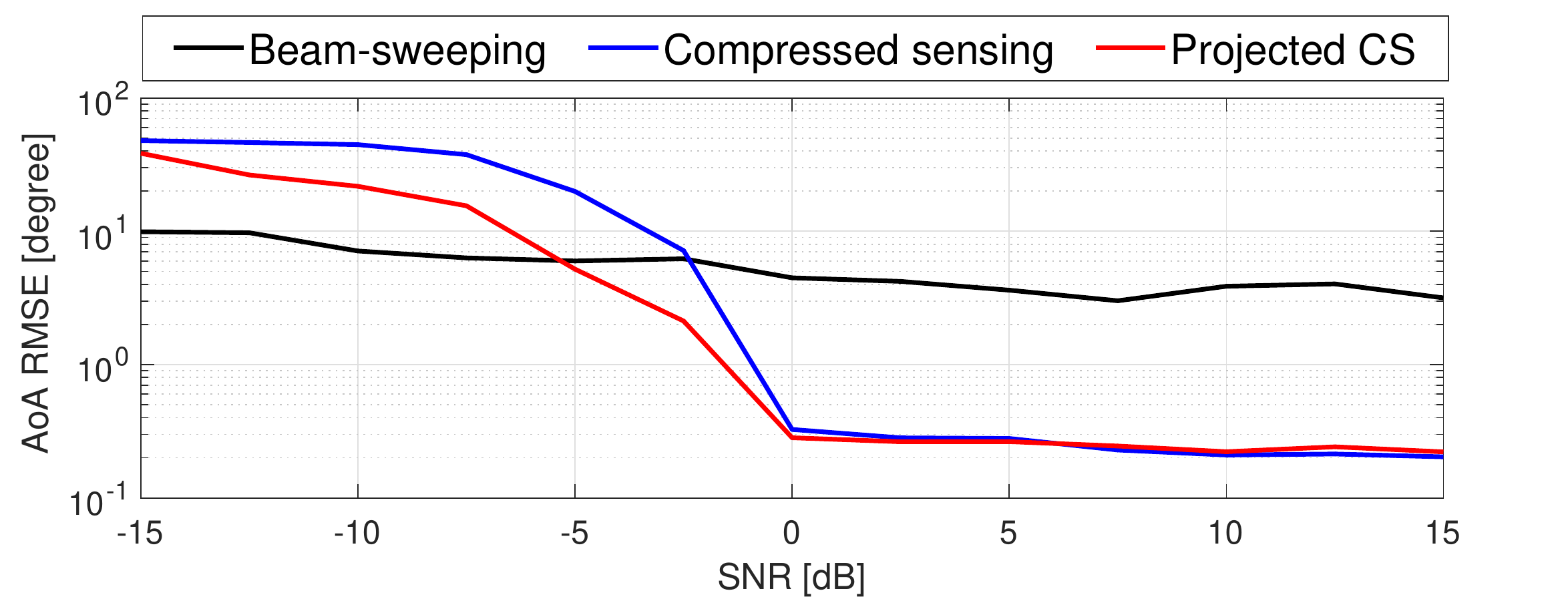}
\vspace{-2mm}
\caption{AoA RMSE of the three algorithms for different SNR levels.}
\vspace{-7mm}
\label{fig:rmse}
\end{center}
\end{figure}

Better tracking triggers of the CS-based and the PCS-based algorithms come from their better AoD/AoA tracking accuracy as compared to that of beam-sweeping. In Fig.~\ref{fig:rmse}, we compare algorithms in terms of AoA tracking accuracy and express the result in terms of RMSE of AoA estimation for 300 Monte Carlo (MC) runs at different SNR levels. The result for the AoD RMSE is similar, and we omit to include it for brevity. The PCS-based algorithm has lower AoA RMSE than the CS-based algorithm in low SNR regime, but both RMSEs experience floor due to finite angular resolution $Q_{\text{MS}}$. Beam-sweeping retains high AoA RMSE regardless of the SNR level due to small number of scanned directions and coarse angular resolution.

\begin{figure}[t]
\begin{tabular}{cc}
\vspace{0mm}
\subfloat[Spectral efficiency with the first model of AoD/AoA changes.]{%
  \includegraphics[clip,width=0.98\columnwidth]{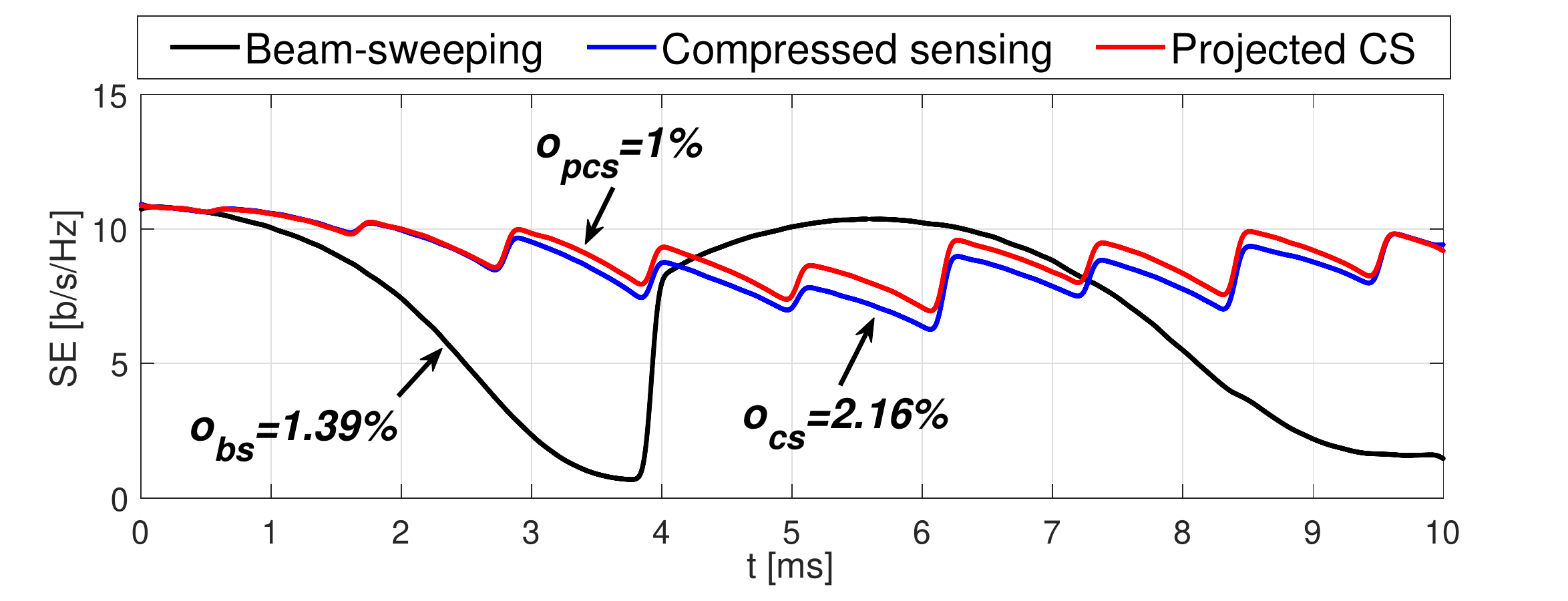}%
}\\
\vspace{0mm}
\subfloat[Spectral efficiency with the second model of AoD/AoA changes.]{%
  \includegraphics[clip,width=0.98\columnwidth]{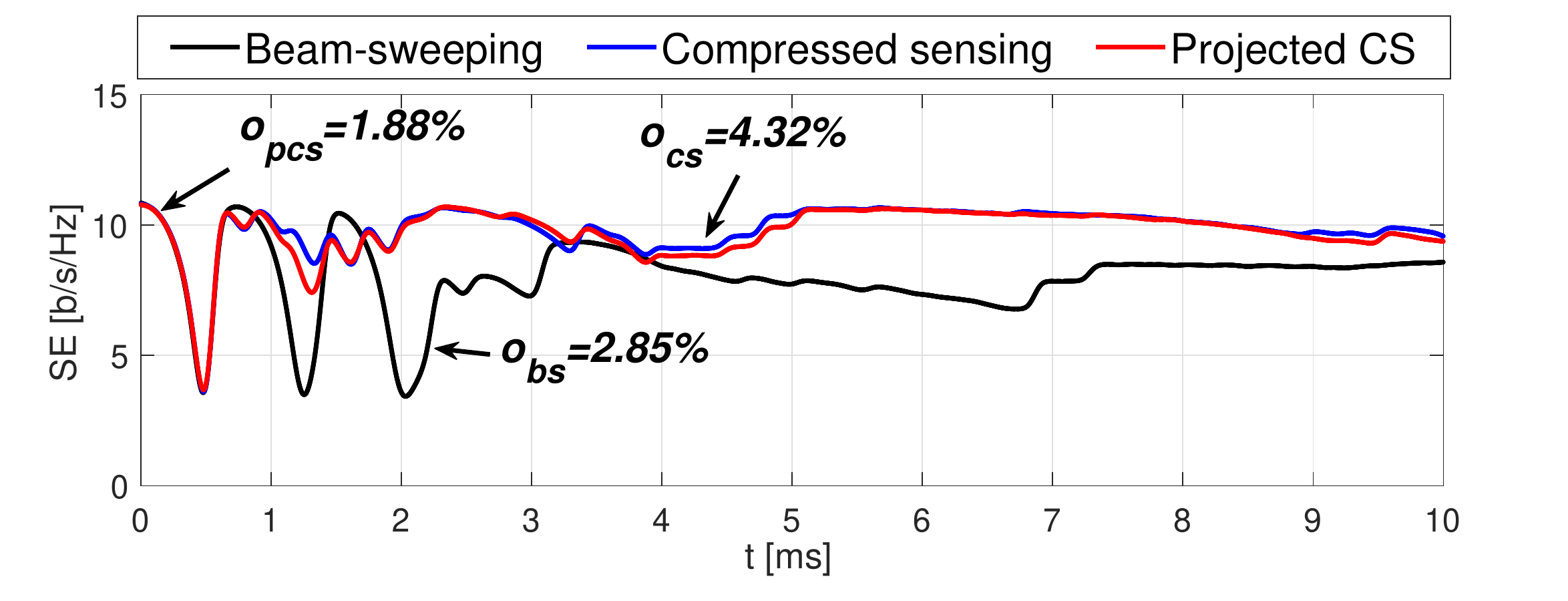}%
}\\
\vspace{-4mm}
\end{tabular}
\caption{Average spectral efficiencies during a period of one time frame.}
\vspace{-5mm}
\label{fig:spectral_vs_time}
\end{figure}

In Fig.~\ref{fig:spectral_vs_time}, we compare spectral efficiencies (SE) of the algorithms as functions of time, averaged over 300 MC runs at $\text{SNR}=0$dB. We also find average required overheads $o_{\text{bs}}$, $o_{\text{cs}}$, and $o_{\text{pcs}}$, when aperiodic tracking is used in conjuction with beam-sweeping, the CS-based algorithm, and the PCS-based algorithm, respectively. We assume that $\hat{\theta}^k=\theta^k=\ang{12}$ and $\hat{\phi}^k=\phi^k=\ang{15}$ for $k=0$, i.e., perfect AoD/AoA estimates are available at the beginning of the frame. With the first model of angular changes, the PCS-based algorithm has higher spectral efficiency than the CS-based algorithm, and both of them are better than beam-sweeping which suffers from coarse angular resolution. Beam-sweeping algorithm waits until the angles significantly change, and then it updates AoD and AoA. Multiple notches in spectral efficiency of the CS-based and the proposed algorithms come from the fact that in a number of MC runs channel tracking periods can be offset, as discussed earlier. The PCS-based and the CS-based algorithms have similar spectral efficiencies when the second model of angular changes is used. Their deep notch at the beginning is a result of choosing $T^1=560$ slots which can be too long for fast angular changes. Nevertheless, both algorithms can recover and adapt $T^t$ for the future channel tracking.  Beam-sweeping is unable to successfully track fast AoD/AoA changes, which results in low spectral efficiency.

\begin{figure}[t]
\begin{center}
\includegraphics[clip,width=0.49\textwidth]{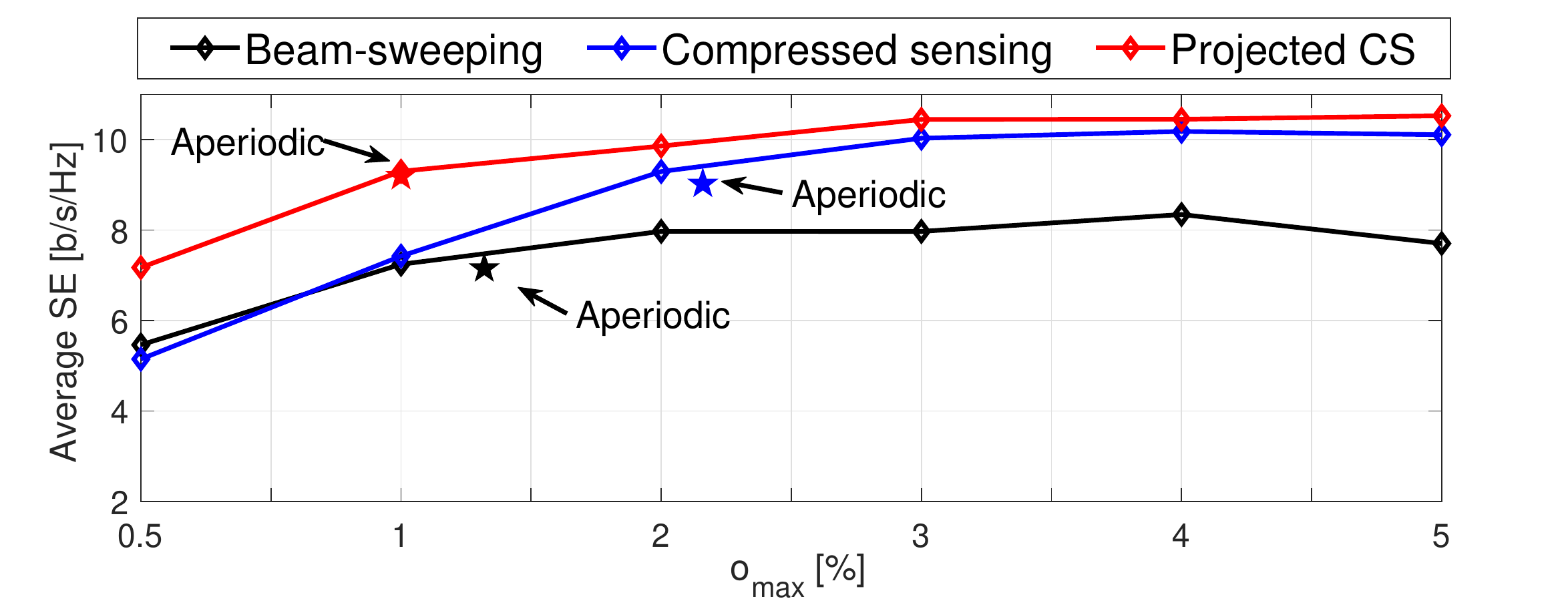}
\vspace{-5mm}
\caption{Average SE within a frame for the first model of AoD/AoA when periodic tracking with different maximum overheads is used.}
\vspace{-5mm}
\label{fig:spectral_vs_overhead}
\vspace{0mm}
\end{center}
\end{figure}

In order to highlight advantage of the proposed algorithm, we compare average spectral efficiencies within a frame at $\text{SNR}=0$dB when periodic tracking is used and maximum overhead is predefined. For given maximum overhead $o_{\text{max}}$, we determine the maximum number of channel tracking periods within a frame for each algorithm, i.e., we find $R_{\text{bs}}$, $R_{\text{cs}}$, and $R_{\text{pcs}}$, and then we choose corresponding tracking periodicities $T_{\text{bs}}$, $T_{\text{cs}}$, and $T_{\text{pcs}}$ from the set $S_T$. We calculate time offset before the first channel tracking as $T^{\text{off}}_{\text{x}}=\ceil[\bigg]{\frac{T-(R_{\text{x}}-1)T_{\text{x}}}{2}}$, where $\text{x}=\{\text{bs},\text{cs},\text{pcs}\}$. In Fig.~\ref{fig:spectral_vs_overhead}, we show results for the first model of AoD/AoA changes and observe that the proposed algorithm achieves the highest average SE regardless of given maximum overhead. The performance with aperiodic tracking tracking is included in the figure, and we observe that aperiodic tracking tends to optimize performance by increasing the average spectral efficiency and decreasing required overhead at the same time.

\begin{figure}[t]
\begin{center}
\includegraphics[clip,width=0.49\textwidth]{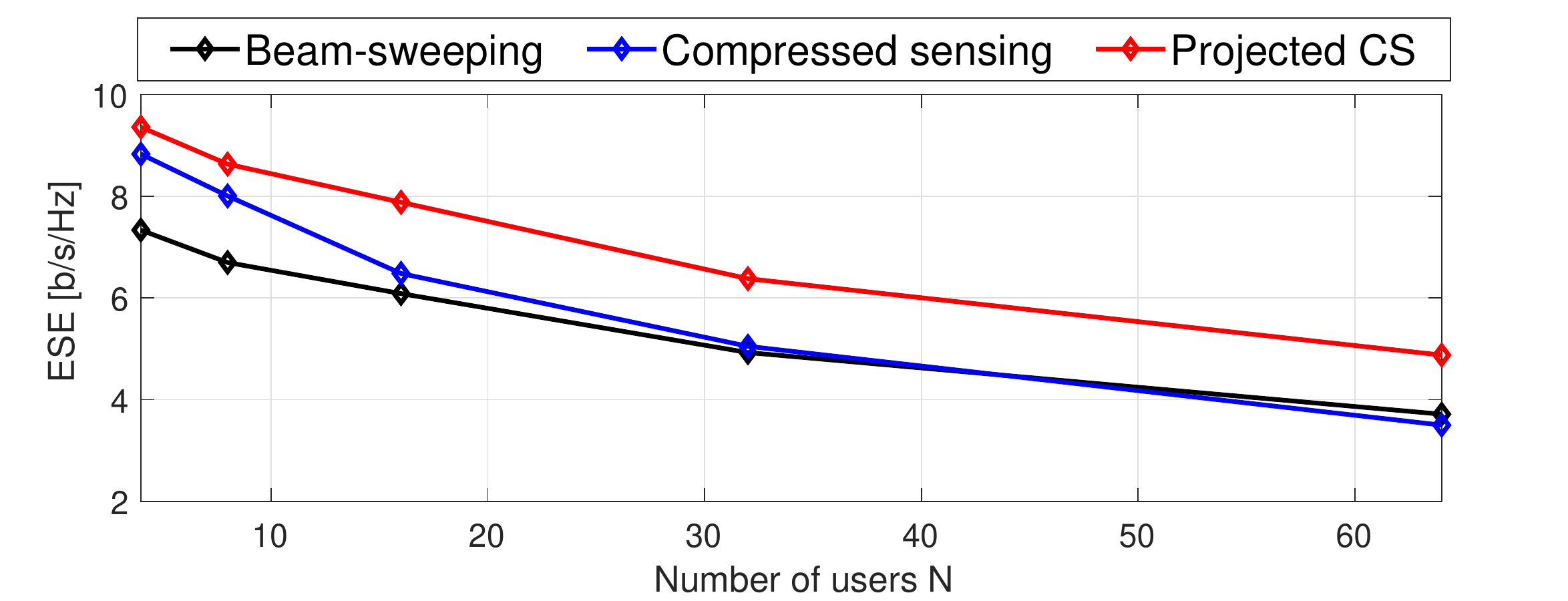}
\vspace{-5mm}
\caption{Average ESE within a frame for the first model of AoD/AoA changes and different number of users $N$.}
\vspace{-7mm}
\label{fig:ESE_vs_users}
\end{center}
\end{figure}

In Fig.~\ref{fig:ESE_vs_users}, we show effective spectral efficiency (ESE) for the first model of AoD/AoA changes, and we note that similar results are obtained for the second model. The ESE is calculated for each algorithm in the following way
\begin{align}
    \text{ESE}(N) = \max_{(\text{SE},o_{\text{max}})}(1-No_{\text{max}})\text{SE},
\end{align}
where $N$ is the number of users and $(\text{SE},o_{\text{max}})$ is a 2-tuple that represent calculated pairs from Fig.~\ref{fig:spectral_vs_overhead}. The PCS-based algorithm achieves the highest ESE regardless of the number of users in the network. It is worth noting that we considered an extreme case when all mobile stations require channel tracking at different time instances and that the CS-based algorithm would perform better if multiple users were tracked at the same time.

%
%
\section{Conclusions}
\label{sec:conclusions}
We proposed a new channel tracking scheme that utilizes knowledge of the previous AoD/AoA estimates to asymmetrically scan beamspace by projecting pseudo-random sequences onto it. A simple implementation of aperiodic channel tracking was proposed and proved to provide high spectral efficiency with low overhead. Our results showed that the PCS-based algorithm is advantageous over beam-sweeping and the CS-based algorithm in terms of AoD/AoA accuracy, achievable spectral efficiency, and required overhead. In the future work, we will design a tracking algorithm that predicts the next AoDs and AoAs and thus significantly reduces the tracking overhead.

%
\bibliographystyle{IEEEtran}
\bibliography{IEEEabrv,references}

\end{document}